# From Static Pathways to Dynamic Mechanisms: A Committor-Based Data-Driven Approach to Chemical Reactions

Radu A. Talmazan [†] and Christophe Chipot [*,†,‡,¶]

†*Laboratoire International Associé Centre National de la Recherche Scientifique et University of Illinois at Urbana-Champaign, Unité Mixte de Recherche n⁰7019, Université de Lorraine, 54506 Vandœuvre-lès-Nancy Cedex, France*
‡*Theoretical and Computational Biophysics Group, Beckman Institute, and Department of Physics, University of Illinois at Urbana-Champaign, Urbana, Illinois 61801, USA*
¶*Department of Biochemistry and Molecular Biology, The University of Chicago, Chicago, Illinois 60637, USA*

E-mail: chipot@uchicago.edu

## Abstract

As computational chemistry methods evolve, dynamic effects have been increasingly recognized to govern chemical reaction pathways in both organic and inorganic systems. Here, we introduce a committor-based workflow that integrates a path-committor-consistent artificial neural network (PCCANN) with an iteratively trained hybrid-DFT-level message passing atomic convolutional encoder (MACE) potential. Beginning with a static nudged elastic band path, PCCANN extracts a committor-consistent string to represent the reactive ensemble. We illustrate the power of this methodology through two representative applications. First, we investigate an $S_NAr$ reaction using MACE trained at hybrid DFT level with implicit solvent. The mechanism is found to be concerted, and the dynamic approach reveals a lower barrier than static treatments. Second, we apply the same protocol to the isomerization of protonated isobutanol to protonated 2-butanol, yielding a quantitatively accurate free-energy landscape. We uncover three competing channels: the established concerted mechanism and two asynchronous stepwise routes mediated by water and methyl transfer, all with comparable activation barriers. Notably, the stepwise pathways traverse metastable intermediates that, to the best of our knowledge, have not been described in prior mechanistic studies. Calculated barrier heights and intermediate stabilities are in close agreement with high-level DFT benchmarks, demonstrating the framework's accuracy. Together, these studies highlight mechanistic diversity across distinct systems and establish the synergistic PCCANN–MACE protocol as a proof-of-concept approach for committor-based discovery of complex reaction dynamics.

## Introduction

Traditional quantum chemical approaches to discovering reaction mechanisms rely heavily on static potential energy surface (PES) analysis, where transition states are located and intrinsic reaction coordinates (IRC) are traced at 0 K. While powerful, these methods often fail to capture the full complexity of chemical reactivity in systems where dynamic effects dominate.[1–4] A growing body of research highlights the limitations of static PES methods in systems exhibiting post-transition state bifurcations (PTSB)—where a single transition state leads to multiple products without intervening intermediates. Such PTSBs complicate kinetic predictions and necessitate a dynamical treatment of reactivity, particularly in heterogeneous environments where anharmonic motion and entropic contributions are substantial. Through these approaches, substantial variability in activation barriers—sometimes exceeding 40–80 kJ mol$^{-1}$ - compared to static estimates have been revealed, further underscoring the importance of including dynamic effects in models.[1,3]

Accurately describing these processes requires molecular dynamics (MD) simulations that incorporate finite-temperature effects, anharmonic motion, and entropic contributions. However, conventional MD is limited by the high energy barriers typical of chemical transformations, which are rarely crossed on accessible timescales. Enhanced sampling methods—such as metadynamics,[5,6] umbrella sampling,[7,8] Blue Moon ensemble sampling,[9] and well-tempered metadynamics with extended-system adaptive biasing force (WTM-eABF)[10,11]—address this challenge by accelerating rare event sampling and enabling exploration of reactive pathways. A persistent difficulty, however, lies in the definition of collective variables (CVs) that faithfully represent the true reaction coordinate in a dynamic system, along with the pathway it follows as the reaction proceeds. While traditional static calculations can obtain a reaction pathway by determining the transition state and following an IRC at 0 K, this does not necessarily represent the true reaction pathway in a dynamic system. Thus, a method to obtain a meaningful, realistic pathway is required to achieve a better understanding of the chemical reaction at hand. A starting point for the discovery of a realistic transition pathway is the committor function, which quantifies the likelihood that a configuration proceeds to the product ensemble before reaching that of the reactant.[12] In the case of MD sampling, by calculating the committor function, we are able to infer more information about the reaction, including the definition of the true dynamic transition state ensemble, represented by configurations with a committor value of 0.5. This probabilistic metric offers a robust alternative to static saddle points and is especially powerful for resolving bifurcated or more complex reaction pathways. It is also worth noting that the committor framework is not restricted, by its very definition, to RRKM-type kinetics. Because committor functions are inherently agnostic to the statistical assumptions of RRKM theory, the approach can, in principle, capture both RRKM and non-RRKM scenarios within a unified description, naturally incorporating dynamical effects such as incomplete energy redistribution and mode-specific energy flow. Furthermore, with the recent introduction of the path-committor-consistent artificial neural network (PCCANN)[13], we are now able to directly obtain a committor consistent string (CCS), representing the transition pathway from reactant to product, in CV space, simultaneously with the committor. This allows us to determine truly dynamic reaction mechanisms and their associated free-energy barriers, directly from MD sampling.

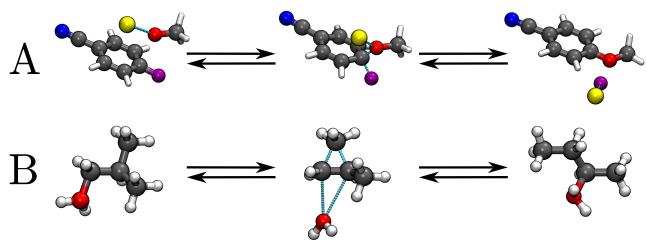

Scheme 1. A. Reaction scheme for the nucleophilic substitution reaction reaction of potassium methoxide and 4-fluorobenzonitrile. B. Reaction scheme for the I$_2$ isomerisation of protonated isobutanol.

To demonstrate the power of combining the aforementioned methods to overcome the limitations of static approaches, we integrate both static and dynamic computational methods to investigate two chemical systems. Firstly, we investigate the bimolecular aromatic nucleophilic substitution reaction ($S_N$Ar) reaction of potassium methoxide and 4-fluorobenzonitrile. This reaction has been investigated previously within an IRC context, to determine the nature of the mechanism. It was observed that different halogens and functional groups attached to the benzyl moiety affected the concerted or stepwise nature of the chemical reaction[14]. The reaction depicted in Scheme 1A was shown to proceed in a concerted fashion. This system repre-sents a good benchmark to confirm the accuracy of the MACE potential when comparing to reference DFT data and it's ability to replicate the concerted mechanism, while including implicit solvent effects.

The second investigated reaction is the $I_2$ isomerization of protonated isobutanol, as illustrated in Scheme 1B, a particularly challenging model system with respect to both kinetics and pathway determination. The isobutanol isomerization and subsequent dehydration to but-1-ene, have been shown to be complex chemical transformations, where static energy profiles have proven to be incomplete, obscuring bifurcations, non-statistical behavior, and competing pathways.[15,16] A density functional theory (DFT) based dynamic study identified several reaction pathways for the conversion of isobutanol to linear butene: a stepwise route involving isomerization to butan-2-ol, followed by E2-type dehydration, and a concerted one-step mechanism combining isomerization and dehydration.[16] These calculations suggest that the stepwise pathway is energetically favored at low temperatures, with the $I_2$ transformation serving as a critical intermediate; at elevated temperatures, however, the concerted mechanism becomes more dominant. A veritable network of possible reaction pathways connects various reactants and products, demonstrating the underlying complexity inherent in such organic chemical reactions.[17]

In our approach, static DFT calculations provide an initial approximation of the energy landscape and transition pathway, while enhanced sampling techniques — specifically WTM-eABF — enable exploration of the reaction dynamics. These simulations initially rely on the GFN2-xTB semi-empirical method, allowing us to discover chemical transition pathways simultaneously with the committor, using PCCANN. The obtained CCSs are then iterated upon, until convergence. The same procedure is repeated, using a MACE-based MLP, trained on the GFN2-xTB sampled geometries, recomputed at a hybrid DFT level, allowing us to achieve higher accuracy. Our findings confirm the concerted nature of the $S_NAr$ reaction and provide mechanistic clarity on isobutanol isomerization, thus establishing a versatile protocol for committor-based reaction discovery in complex chemical systems.

## Methods

### Static calculations and benchmarking

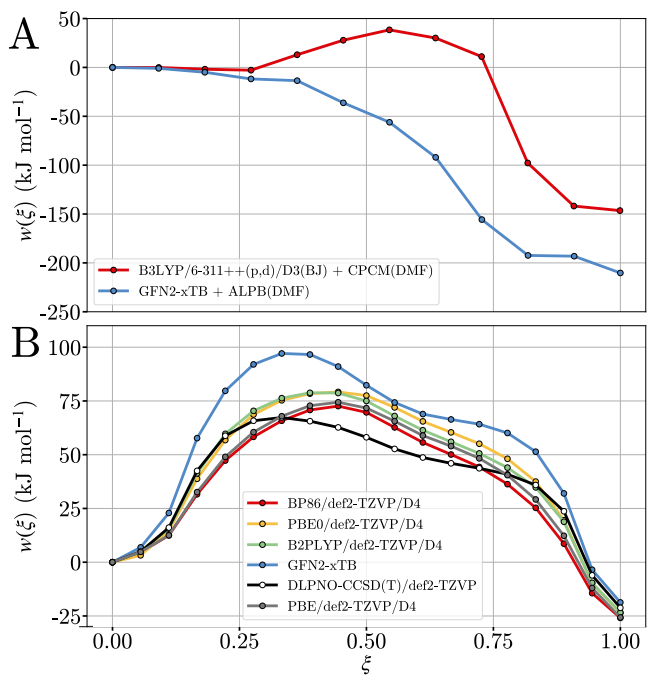

Figure 1. A. Free energy profile of the $S_N$Ar reaction, computed at a static level, with the rigid rotor harmonic approximation used for estimating the thermal and entropic corrections at 300 K. B. Free energy profile of the $I_2$ isomerization reaction, computed at a static level, with the rigid rotor harmonic approximation used for estimating the thermal and entropic corrections at 300 K.

For the $S_N$Ar reaction, we did not undertake a new evaluation of DFT functional performance, as a comprehensive benchmark had already been provided in the reference publication.[14] Based on those results, we employed the B3LYP functional[18,19] with D3-BJ dispersion corrections[20], and the 6-311++(p,d) basis set,[21,22] which showed low errors for this system. Prior work further demonstrated that including the cation in the chemical model enables implicit solvation to capture all relevant effects, whereas explicit solvation only adds complexity without improving accuracy.[14] Accordingly, solvent effects were mod-eled using the conductor-like polarizable continuum model (CPCM)[23] in combination with DFT, and the analytical linearized Poisson-Boltzmann (ALPB) model[24] in the case of GFN2-xTB. In our workflow, GFN2-xTB was used to generate the initial dataset for MACE MLP training. The semi-empirical method fails to reproduce the correct energetic profile, showing no discernible barrier, while B3LYP predicts a free-energy barrier of 39 kJ mol$^{-1}$. Similarly, GFN2-xTB substantially overestimates the free-energy release associated with the formation of the product by about 70 kJ mol$^{-1}$.

In the case of the $I_2$ isomerization, we tested several DFT functionals, namely BP86,[25,26] PBE,[27] B3LYP,[18,19] PBE0,[28] and B2PLYP[29] for performance and accuracy, alongside GFN2-xTB.[30]. Each functional was tested using the def2-TZVP basis set.[31] As the reference, DLPNO-CCSD(T)[32]/def2-TZVP[31] calculations were performed on the PBE geometries. All calculations were performed with D4 dispersion corrections, where applicable.[33]

For both systems, the initial reaction mechanisms were determined by calculating a static profile using the NEB approach.[34,35] This allowed us to obtain a basic profile of the reaction and identify the corresponding transition state.

For the protonated isobutanol system, the static calculations revealed that while all methods approximate the DLPNO-CCSD(T)/def2-TZVP//PBE/def2-TZVP/D4 reaction energy correctly, there are significant variations in the transition area along the pathway, as shown in Figure 1. While GFN2-xTB overestimates the energy barrier significantly, it, nonetheless, remarkably maintains a qualitatively similar reaction profile, while the geometries of the structures are nearly identical to those obtained with DFT. From the various DFT functionals, we observe similar performance and accuracy from PBE and BP86. Likewise PBE0 and B2PLYP deliver quantitatively similar energy profiles. Notably, PBE0 and B2PLYP match the CCSD(T) reference energies remarkably well, as the structures approach minima on the static free-energy surface, while maintaining the same general qualitative curve shape. Thus, we opted for the PBE0 functional, in conjunction with def2-TZVP and D4 dispersion corrections, as our DFT reference and therefore our base for training the MACE model. See Supporting Information for full details.

## Sampling with WTM-eABF

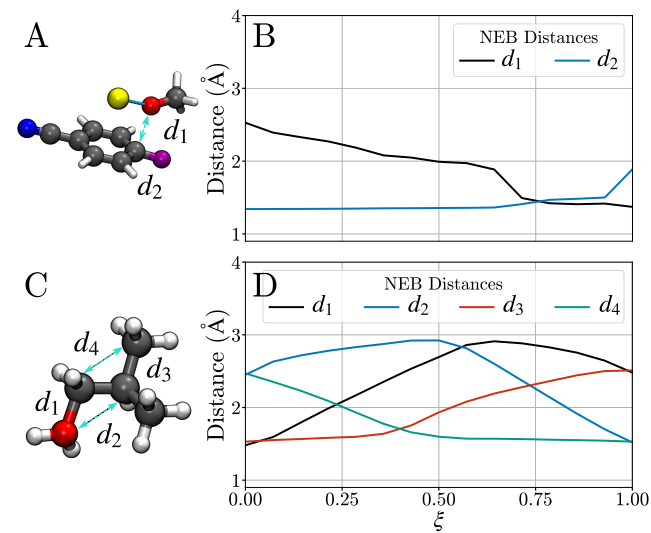

Figure 2. A. The two distances defining the CV, namely the carbon-oxygen distance ($d_1$) and the carbon-fluoride distance ($d_2$). B. The behavior of these two distances along the minimum energy pathway obtained from the NEB. C. the four distances defining the CV, namely the two carbon-oxygen distances ($d_1$ and $d_3$) and the two carbon-carbon distances ($d_2$ and $d_4$). D. The behavior of these four distances along the minimum energy pathway obtained from the NEB.

To overcome the substantial energy barriers associated with the chemical reactions, we employed WTM-eABF, which combines efficient barrier crossing with accurate free-energy estimation.[36–38] This approach facilitates efficient exploration of reaction pathways and reliable identification of intermediates and transition states, overcoming the sampling limitations of conventional MD. We used this method in combination with path-collective variable (PCV) sampling, projecting molecular transitions onto a low-dimensional path, allowing enhanced exploration of free-energy landscapes and accurate identification of reaction pathways.[39–41]

For the $S_NAr$ reaction, the CV was defined by the two bonds central to the transformation: the carbon–fluoride bond that breaks to release fluoride as the leaving group, and the carbon–oxygen bond that forms as the methoxy group attaches to the benzene ring (see Figure 2A). The evolution of these coordinates along the minimum energy pathway obtained from the NEB calculation is shown in Figure 2B.

For the isobutanol example, the CV was defined by four interatomic distances, shown in Figure 2C and D, corresponding to the geometry changes from reactant to product, using the arithmetic CV implemented in the Colvars library,[42,43]. In both cases, the initial sampling along the NEB-derived PCV was used to train PCCANN, allowing us to identify distinct pathways. Each pathway was iteratively refined through additional WTM-eABF sampling, until the committor-consistent string (CCS) stabilized (see Supporting Information for details on convergence). The same protocol was applied for both GFN2-xTB and MACE sampling. Full parameter details for the WTM-eABF simulations are provided in the Supporting Information.

## Machine-learning potential

As direct DFT-level sampling of the systems was computationally prohibitive — even with a GGA functional such as BP86 — we employed a custom-trained MACE potentials to achieve hybrid DFT accuracy at a tractable cost. Benchmarking of the pretrained MACE foundational models provided with the MACE release (v0.3.10) revealed insufficient accuracy, likely due to the complex non-covalent interactions in the $S_NAr$ system and the net +1 charge in the case of protonated isobutanol. The custom MACE potential was trained via an iterative active-learning protocol to reproduce hybrid DFT energetics and forces (B3LYP/6-311++(d,p)/D3(BJ) + CPCM(DMF) for $S_NAr$ and PBE0/def2-TZVP/D4 for protonated isobutanol). The initial models were trained on structurally diverse sets of configurations obtained from GFN2-xTB WTM-eABF sampling. The subsequent iterations alternated between WTM-eABF sampling, error analysis against DFT reference energies, and targeted augmentation of under-represented regions — particularly near the transition state, where initial errors approached 25 kJ mol$^{-1}$. Incorporating additional transition structures along the pathway reduced these errors to the single kJ mol$^{-1}$ range. Final production sampling included on-the-fly DFT evaluation of 1% of configurations to monitor model fidelity (see SI for full training protocol and parameters).

## Clustering of trajectories

We analyzed the molecular dynamics trajectories using k-means clustering to identify representative intermediate and transition state structures. The optimal number of clusters for each ensemble was determined by evaluating the Silhouette Score,[44] Calinski-Harabasz index,[45] and Davies-Bouldin index,[46] as well as considering the distribution of cluster populations and the average distance to cluster centroids (see Supporting Information Figure S8 and Figure S9). Clustering was performed on the basis of the PMF profile, with transition states assigned to the regions corresponding to the free-energy maxima and intermediate states selected from the wells between these maxima. The representative intermediate structures from each cluster were subsequently optimized using DFT to assess whether they correspond to true minima on the potential energy surface.

## Technical implementation

To determine the committor and corresponding reaction pathway, we adopted the iterative strategy introduced in the original PCCANN study[13], illustrated schematically in Figure 3A. The procedure began with an initial sampling along the pathway obtained via the NEB method, followed by the determination of the committor and associated CCS. Additional sampling was generated along this CCS, and the cycle was repeated until the CCS RMSD relative to the preceding string fell below 0.05 Å. An analogous iterative protocol was applied during the training of the MACE MLP potential (see Figure 3B), as described in the Machine-learning potential section above. The full technical implementation is freely available at github.com/Lapsis-glitch/Data_driven_path_discovery.

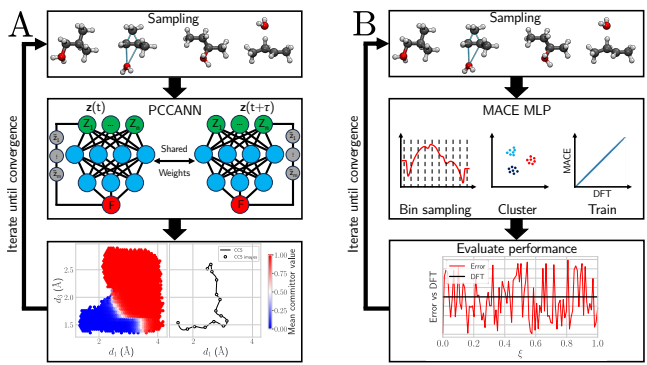
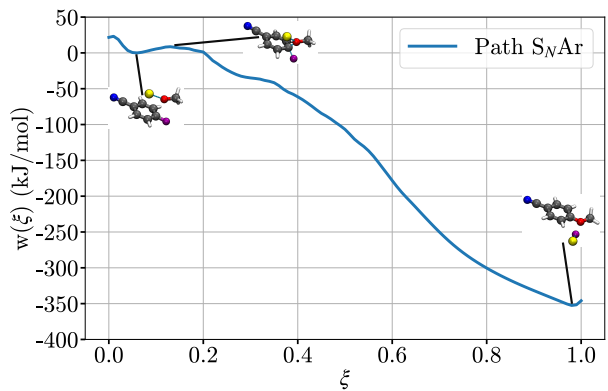

Figure 3. A. Overview of the PCCANN procedure for obtaining a converged committor and corresponding committor consistent string. B. Workflow for the training of the MACE MLP through an iterative loop, employing enhanced-sampling methods and repeated refinement of the MLP.

Figure 4. Free energy profile along the reaction coordinate model, resulting from the GFN2-xTB sampling of the reaction pathway obtained via clustering of the committor at 300 K.

All quantum-chemical calculations were performed with ORCA 6.0.1.[47] MD sampling was obtained by leveraging NAMD 3.0.1,[48] in conjunction with the Colvars library module.[42,43] GFN2-xTB version 6.7.0 was used. MACE version 0.3.10 was used. The system topologies for the MD simulations were generated with PyConSolv 1.0.6.[49] All figures and plots were generated with Python 3[50] and VMD 2.0.[51]

# Results and Discussion

## GFN2-xTB dynamics

### $S_N$Ar reaction

The GFN2-xTB method yielded reasonable system geometries, though it tended to overestimate the free-energy decrease associated with the reaction. As shown in Figure 4, the resulting energy profile contains no intermediate structure, in agreement with previous DFT-based studies of the system. In contrast to the static calculations, however, a free-energy barrier of 8.7 kJ mol$^{-1}$ is observed.

### $I_2$ Isomerization of isobutanol

GFN2-xTB facilitates extensive sampling, enabling a first characterization of the reaction profile, committor evaluation, and identification of competing mechanistic pathways. Initial sampling along the NEB-derived path yielded a free-energy profile (see Figure 5) that qualitatively agrees with previous computational studies,[16] albeit with a systematically elevated barrier, consistent with the known GFN2-xTB behavior, as exhibited in benchmarking (see Figure 1). Committor analysis and clustering revealed three distinct mechanistic pathways (see Supporting Information for the exact protocol). Path 1 corresponds to a concerted mechanism in which water departure and methyl migration occur synchronously, consistent with prior reports.[16] Paths 2 and 3 proceed via stepwise mechanisms, initiated either by water loss (INT$_2$) or methyl shift (INT$_3$), each forming a metastable carbocation intermediate.

All three pathways exhibit comparable barrier heights. Notably, the GFN2-xTB free-energy profiles reveal an additional shallow intermediate, structurally resembling linear butene with a hydrogen-bonded water molecule, with the proton shared between two adjacent carbon atoms (see Figure S23A). Upon optimization at both PBE0/def2-TZVP/D4 and DLPNO-CCSD(T)/def2-TZVP levels, this species converges to a hydrogen-bridged butyl ion, previously reported in high-level ab-initio studies.[52] This intermediate serves as a potential branching node for an alternative dehydration channel leading to the formation of trans-but-2-ene, instead of isobutanol. The final committor-consistent strings for all three pathways converged within three PCCANN iterations.

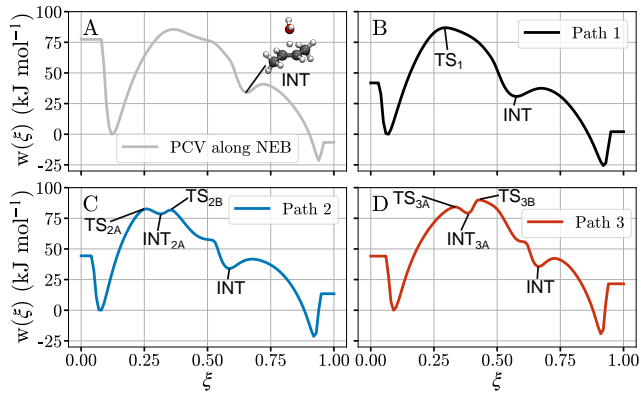
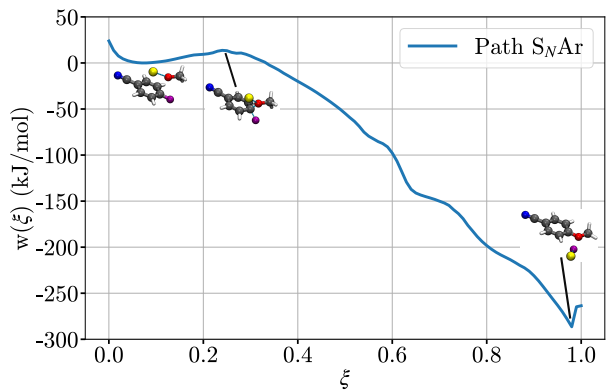

Figure 5. Free energy profiles along the reaction coordinate model, resulting from the GFN2-xTB sampling of the reaction pathways at 300 K: A depicts the free-energy profile along the statically determined NEB transition path; B, C and D depict the free-energy profile along the pathways obtained via clustering of the committor.

Figure 6. Free energy profile along the reaction coordinate model, resulting from the MACE sampling of the reaction pathways at 300 K.

## MACE dynamics

### $S_NAr$ reaction

Compared to GFN2-xTB sampling, MACE revealed a smaller free-energy difference between reactants and products (-286 kJ mol$^{-1}$), consistent with expectations from static computations. Under dynamic conditions, the free-energy barrier is markedly reduced to 13.7 kJ mol$^{-1}$. The resulting pathway closely resembles that obtained from NEB calculations. Clustering of the MACE trajectories further confirmed the presence of a single concerted mechanism, in which the Meisenheimer complex serves as a transition state rather than a stable intermediate (Figure 6).

### $I_2$ Isomerization of isobutanol

Sampling performed with MACE as a replacement for the GFN2-xTB back-end yielded similar mechanistic pathways (Figure 7), including both concerted and stepwise mechanisms. Notably, the activation barrier along the NEB-derived path is considerably lower than that obtained with GFN2-xTB, consistent with expectations from static benchmarking, and compares favorably with previous studies.[16] At elevated temperatures (e.g., 500 K), reaction barriers decrease by approximately 8 kJ mol$^{-1}$ across all pathways, accompanied by changes in the features of the free-energy profiles.

Sampling the isomerization transition pathways at 500 K versus 300 K (Figure 7) reveals the emergence of a shallow post-transition-state intermediate, more prominently observed under elevated thermal conditions. Upon investigation with DFT, this intermediate, characterized by partial charge delocalization and a stretched hydrogen bond, matches that highlighted in the GFN2-xTB simulations. In the high-temperature WTM-eABF trajectories, both concerted and stepwise channels funnel into this species. This intermediate could serve as a branching node: it may proceed through the $I_2$ pathway to form 2-butanol, or divert into a dehydration-like channel yielding trans-but-2-ene. The presence of this bifurcation highlights how increased kinetic energy enables the system to sample adjacent saddle points and alternate product funnels, reshaping the mechanistic landscape. We investigate this further in the reaction mechanism determination section below.

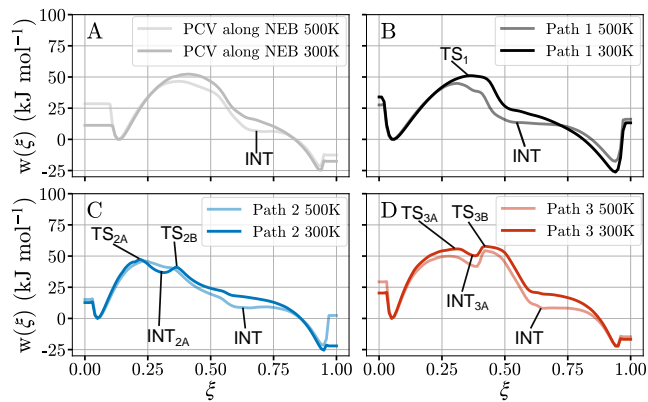

Figure 7. Free-energy profiles along the reaction coordinate model, resulting from the MACE sampling of the reaction pathways at 300 K and 500 K: A depicts the free-energy profile along the statically determined transition path; B, C and D depict the free-energy profile along the pathways obtained via clustering of the committor. The free-energy profiles are taken from the final sampling along the respective converged CCS.

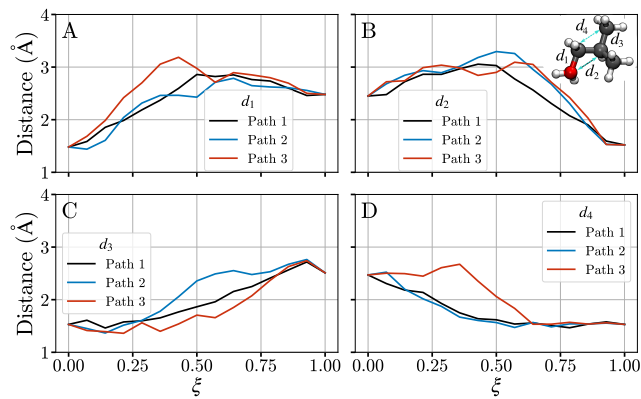

Figure 8. A,B,C and D show the evolution of the four distances defining the CCS, for each discrete pathway, at 300K. In panel B, an inset is shown, detailing which degree of freedom each of the four distances represents.

Examining the four distances defining the CCS, depicted in Figure 8, we observe that the variables defining the pathway diverge from each other at various points along the reaction coordinate. While the latter part of the pathway is common for $d_1$ and $d_4$, the beginning of the transition pathway appears similar in the case of $d_2$ and $d_3$. By computing the pairwise RMSD between the path-ways, shown in the Supporting Information Figure S3, we observe that the differences indeed appear mostly in the transition state regions, as the deviations decrease towards the reactant and product.

The computation of the committor using PC-CANN provides us with another important element, namely the ability to compute reaction rates. By evaluating the flux of the reaction in the direction from reactant to product, we are able to obtain rates of $1.4 \times 10^{-6}$ ps$^{-1}$ to $1.7 \times 10^{-5}$ ps$^{-1}$ at 300 K (see Figure S20) and between $1.6 \times 10^{-5}$ ps$^{-1}$ to $4.6 \times 10^{-5}$ ps$^{-1}$ at 500K (see Figure S21). As there is no experimental estimate of rates for the $I_2$ transformation, we can refer to computationally determined values, via the Eyring equation, using values from 16. The numbers we obtain quantitatively match the computational results at 300K, while at 500K the computed rate is lower (see SI for more details on computing the reaction rates). This decreased rate estimate at 500K can likely be attributed to the energy differences between the PBE and PBE0 functionals.

## Reaction mechanism determination

### $S_N$Ar reaction

In both the GFN2-xTB and MACE sampling approaches, the reaction mechanism is concerted, with the Meisenheimer complex functioning as the transition state rather than a stable intermediate. Static DFT evaluations predict a substantially higher free-energy barrier, whereas the dynamic ensemble approach yields a much lower value, underscoring the influence of dynamical effects on the reaction profile. Moreover, the dynamic treatment exposes the limitations of rigid-rotor harmonic oscillator models in describing non-covalently bound systems. By calculating the free energy within the WTM-eABF framework, entropy contributions from loosely bound molecules associated with another species can be estimated more accurately.

## I$_2$ Isomerization of isobutanol

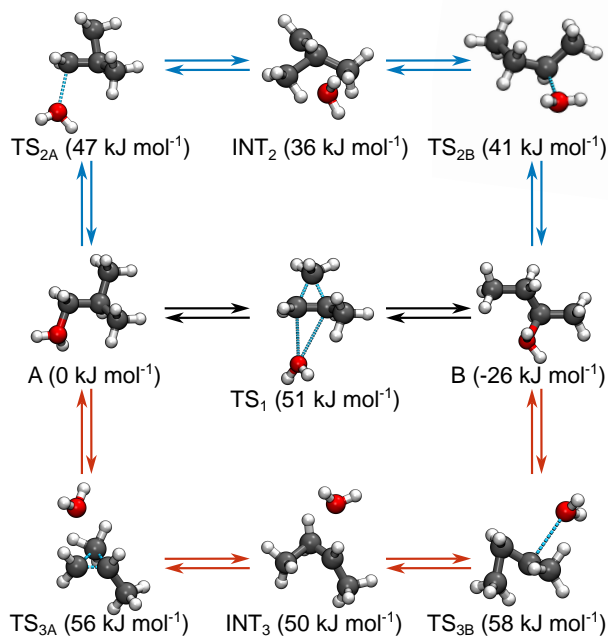

Figure 9. Competing pathways for the isomerization of isobutanol, with corresponding free energy values, relative to the reactant state A. The concerted mechanism, color-coded in black, corresponds to Path 1, the upper pathway, color-coded in blue corresponds to Path 2, while the bottom pathway, in red, corresponds to Path 3.

Both GFN2-xTB and MACE sampling revealed a trifurcation in the reaction landscape for the I$_2$ isomerization of isobutanol to 2-butanol. This three-way split in the pathway emerges as the mechanisms proceed via alternative routes leading up to the transition state along the reaction coordinate (RC). The existence of parallel pathways underscores the ruggedness of the transition-state region, with greater complexity and branching than previously reported. Switching between reaction channels is observed both before and after the transition state, as trajectories naturally converge toward the reactant and product basins (see Supporting Information, Figure S3 and Figure S6). Critically, the coexistence of concerted and stepwise mechanisms highlights how subtle changes in atomic coordination steer the reaction toward distinct routes (see Figure 9). In the concerted branch (Path 1), methyl transfer and C–O bond cleavage occur synchronously in a single transition event. In the stepwise branches (Paths 2 and 3), discrete water-stabilized carbocation intermediates form—INT$_2$ in the water-loss-initiated pathway, and INT$_3$ in the methyl-shift-initiated pathway. Both pathways emerge naturally from the committor-consistent string, but would collapse onto a single "averaged" path in static treatments. Only by allowing the system to explore off-string orthogonal degrees of freedom does the full ensemble of reactive channels become apparent. Notably, the intermediate structures found in the stepwise mechanisms are stable under full DFT optimization (Figure S23), lending further credence to their role as genuine mechanistic waypoints.

While we restricted ourselves to sampling the I$_2$ pathway, based on the mechanistic findings, we can propose additional pathways leading to the formation of secondary products via dehydration reactions. Beyond the three pathways leading to 2-butanol, the intermediate structures in Figure 9 can be considered branching points that are geometrically and electronically predisposed to alternate dehydration products. From both INT$_2$ and INT$_3$, a DH-type elimination can lead to to trans- or cis-but-2-ene, or isobutene. We performed static calculations for these dehydration reactions, which revealed free-energy barriers varying between 6 kJ mol$^{-1}$ and 17 kJ mol$^{-1}$ for the transformations into linear butenes (see Supporting Information for details). The conversion of INT$_2$ to isobutene is however unfavorable, with an electronic energy barrier of over 100 kJ mol$^{-1}$. For the DH-elimination of water from the stabilized post-transition intermediate INT, more prominent at 500 K, we computed free-energy barriers of 34.2 kJ mol$^{-1}$ and 28.8 kJ mol$^{-1}$ to yield trans- and cis-but-2-ene. The preliminary energetic estimates indicate that these dehydration steps are likely competitive with the completion of the I$_2$ pathway, suggesting a bifurcation in the reaction coordinate downstream of the transition state. This highlights the multifaceted nature of the dehydration–isomerization network and the potential for temperature-dependent product selectivity. A full kinetic and thermodynamic evaluation of these alternate channels, integrated into the dynamic pathway analysis, will be the subject of future work. The insights gained from the evaluation of the I$_2$ isomerization underscore that dynamic transition-path finding is not merely a re-finement over static approaches, but a necessity for uncovering hidden mechanistic landscapes when energy barriers are comparable.

# Conclusion

The application of the committor-based workflow to the $S_NAr$ reaction demonstrates that the mechanism proceeds in a concerted fashion. Importantly, the dynamic approach reveals a lower effective barrier compared to static treatments, highlighting how committor-guided sampling can uncover simplified yet more accurate mechanistic pictures even in well-studied transformations. The dynamic, committor-based exploration of the isobutanol $I_2$ isomerization reveals a highly rugged transition-state region with bifurcations emerging before the barrier summit, and frequent channel switching after the transition state. Clustering of sampled metastable configurations identified stepwise intermediates that, upon full PBE0/def2-SVP/D4 geometry optimization and frequency analysis, proved to be genuine minima. This mechanistic complexity far exceeds the single post-transition bifurcation previously reported. In addition to resolving the coexistence of concerted methyl-transfer/C–O-cleavage pathways and stepwise carbocation-mediated channels, our analysis identifies branching points at $INT_2$ and $INT_3$ that are predisposed to alternate dehydration products — including trans-/cis-but-2-ene — with preliminary energetics suggesting kinetic competitiveness. These channels, revealed only through dynamic pathway sampling, underscore the multifaceted nature of the dehydration–isomerization network and the potential for temperature-dependent product selectivity. The synergistic integration of PCCANN with a hybrid DFT-level, iteratively trained MACE potential establishes a powerful framework for dynamic pathfinding in complex chemical systems. By relying on the underlying committor to describe the chemical reaction in the RC space, this framework uncovers parallel, low-energy funnels that static treatments inherently miss. Extending this active-learning-facilitated, committor-based paradigm to larger organic rearrangements, catalytic processes, and enzymatic transformations will be key to predicting both selectivity and kinetics under realistic conditions.

■ ASSOCIATED CONTENT

## Supporting Information

The supporting information contains more details on the methodology regarding the training of the MACE MLP, as well as a more detailed committor and trajectory analysis, combined with the determination of reaction rates and free-energy values.

## Data Availability Statement

The input files for the MD simulations and associated CCS pathways are available on github.com/Lapsis-glitch/Data_driven_path_discovery Completed trajectories are available upon request. The repository likewise contains python implementations of the protocols highlighted in this work, including clustering, error evaluation methods and MACE training scripts.

■ ACKNOWLEDGMENTS

C.C. acknowledges the European Research Council (grant project 101097272 "MilliInMicro"), the Université de Lorraine through its Lorraine Université d'Excellence initiative, and the Région Grand-Est (project "Respire"), and the Agence Nationale de la Recherche under France 2030 (contract ANR-22-PEBB-0009) for support in the context of the MAMABIO project (B-BEST PEPR) for their support.
The authors thank Dr. Celine Chizallet for helpful comments and insights on the manuscript.

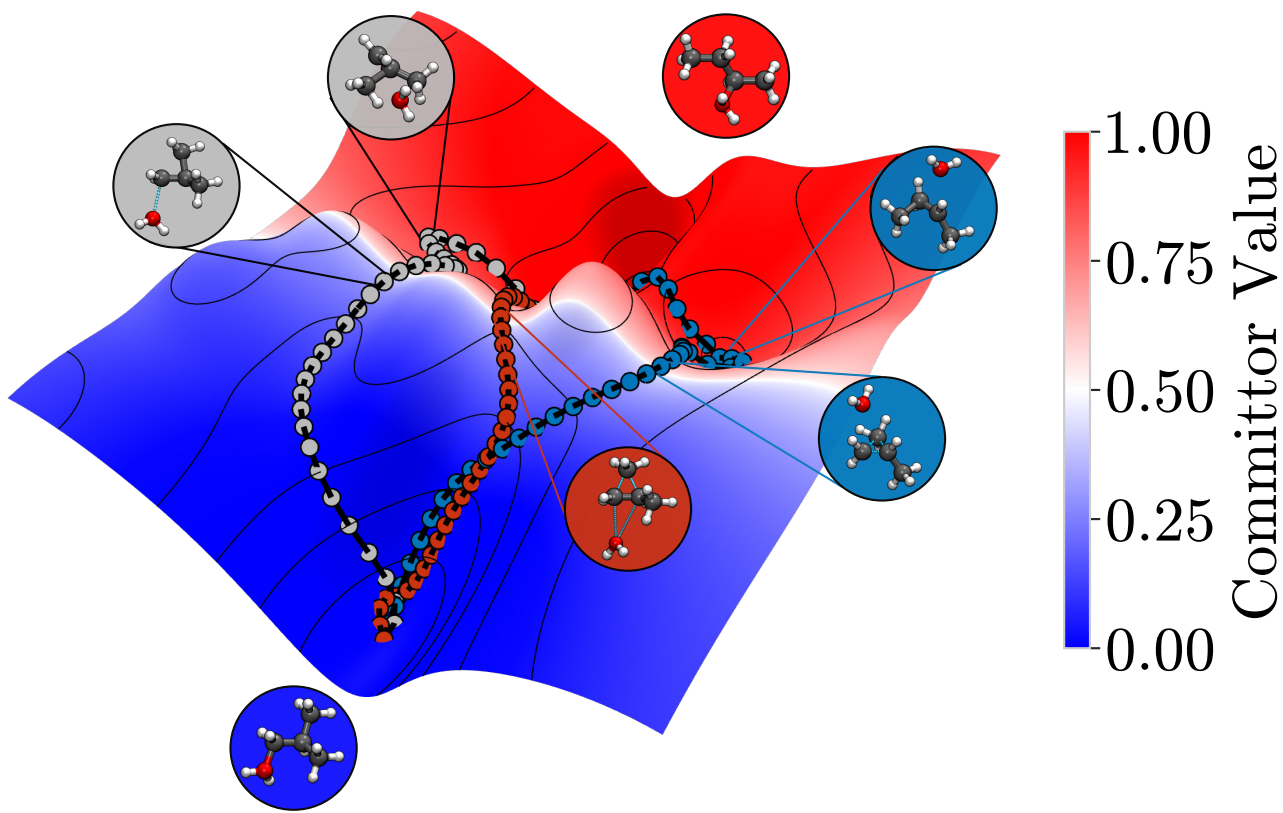

TOC graph